\documentstyle[12pt,fleqn]{article}
\author{L.~Didukh, O.~Kramar and Yu.~Skorenkyy \\
{\small \it Ternopil State Technical University, Department of Physics}
\protect\\ {\small \it 56 Rus'ka Str., Ternopil UA--46001, Ukraine}\\
{\small \it E-mail: didukh@tu.edu.te.ua}}
\date{}
\title{Ground state energy of metallic ferromagnet in a generalized Hubbard model}		
\begin{document}
\maketitle
\begin{abstract}
In the present work ferromagnetic ordering in the Hubbard 
model generalized by taking into account the inter-atomic exchange 
interaction and correlated hopping 
in partially filled narrow band is considered.
In the case of weak electron-electron interaction the  ground state energy
and condition of ferromagnetic state realization are 
found by using the Green function technique. 
The obtained results indicate the important role of correlated hopping.

PACS number(s): 71.10.Fd, 71.30.+h, 71.27.+a
\end{abstract}
\section{Introduction}
The problem of the origin of the metallic
ferromagnetism, in spite of the variety of theoretical attempts to solve it,
still remains open. 
The simplest model at the basis of which one has tried to 
explain the physics of ferromagnetism in $3d$ transition metals is the 
Hubbard model that describes itinerant electrons in a single non-degenerate
band interacting via on-site Coulomb repulsion~\cite{hub}-\cite{gutz}.
The interplay between the kinetic energy of electrons and Coulomb
interaction strongly influences the magnetic ordering. 
In some approximations the ferromagnetic solution in this model was obtained 
for large values of intra-atomic Coulomb interaction $U$; 
in this connection note the Nagaoka's theorem~\cite{nag} (the ferromagnetism 
in half-filled band with single electron or single hole at $U\to \infty$).
The problem of metallic ferromagnetism is studied intensively during last 
years (for recent reviews see Ref. \cite{tas, dieter})
however the undivided opinion is not reached so far.

Nowadays we can distinguish a few ways to obtain the ferromagnetic solution: 
first is
to stay in the framework of the Hubbard model but to use a special lattice 
geometry and density of states; second is to go beyond the Hubbard model 
including the band degeneracy (note that real ferromagnetic
materials have the orbital degeneracy); and third is to take into account 
other matrix elements of electron-electron interactions in addition
to the intra-atomic Coulomb interaction. We believe that third way can 
give the qualitative  explanation of the magnetic 
properties of narrow band materials although the construction of consistent 
theory of metallic ferromagnetism should include the orbital degeneracy
and real density of states. The aim of this article is to study the influence 
of so-called "off-diagonal" matrix elements of electron-electron interaction 
on the condition of ferromagnetism stabilization. In spite of the fact 
that the magnitudes of the neglected in the Hubbard Hamiltonian
terms  are small in comparison with on-site repulsion or band hopping 
integral but they can essentially influence
(assist or suppress) the ferromagnetic ordering. The importance of some 
of these matrix elements was pointed out in many works \cite{did}-\cite{camp};
here we note the special role of the direct exchange interaction 
and correlated hopping (taking into account of the inter-atomic
density-density Coulomb interaction which play essential role in charge
ordering goes beyond the goal of this article).

Last ten years
the problem of importance of exchange interaction again is topical 
especially in the works~\cite{hir_t0}-\cite{amad},
where it was concluded that the nearest-neighbour exchange 
interaction  $J$ plays a fundamental role for the existence of ferromagnetism. 
In the cited papers the condition for ferromagnetism  in the framework of 
mean-field theory has been obtained; also the exact diagonalization solutions 
for one-dimensional system with different band-fillings have been constructed 
and compared with results of mean-field theory \cite{amad}.

The exact criteria for the arising of ferromagnetism in the 
partial case of the model with correlated hopping have been
derived by Strack and Vollhardt \cite{Str&Voll}. These criteria also 
show the significance of the direct exchange interaction.      
Vollhardt and co-workers \cite{voll,wahl} within dynamical mean-field theory
with the asymmetrical density of states having a peak near the band edge
pointed out that already small direct exchange can stabilize the
ferromagnetic ordering. 

As mentioned there is an additional mechanism which can 
favour ferromagnetic ordering. The "off-diagonal" matrix elements
of Coulomb interaction of electrons include the term that describes the 
density-dependent hopping -- so-called correlated hopping. The importance of 
correlated hopping and its considerable place in the descriptions of 
metal-insulator transition and superconductivity are known \cite{cmp}-\cite{bs}.
So we expect that taking into account the correlated hopping can give more 
clear understanding of the metallic ferromagnetism. The attempt
to take into account the influence of correlated hopping has been done by 
Hirsch in  Ref.~\cite{amad}. It has been  shown that correlated
hopping can give rise to the asymmetry of the condition of ferromagnetic
ordering
(to enhance the tendency to ferromagnetism of electron-like 
versus hole-like carriers for small $U$, and have the opposite effect for 
large $U$). In our previous work we also have pointed out the importance
of electron-hole asymmetry in the theory of metal-insulator transition and 
metallic ferromagnetism~\cite{cmp}. It is important that correlated hopping
give rise to concentration dependency of hopping integral.

The purpose of this paper is to investigate the possibility
of ferromagnetic ordering and to study both the influence
of correlated hopping of electrons and direct exchange
interaction on the condition of ferromagnetism. The paper is organized
as follows. In the Section 2 the Hamiltonian of the model with electron-hole
asymmetry is formulated. In the Section 3 the single-particle Green function 
and energy spectrum are found in the case of weak electron-electron 
interaction. In the Section 4 the ground state energy of the model and 
the criteria of ferromagnetism stabilization are calculated
and discussed. Section 5 is devoted to the conclusions. 

\section {The model Hamiltonian}
In accordance with the works~\cite{did,jps1,cmp} we take into account 
in the Hamiltonian the matrix elements $J(ijji)=J$, $J(ikjk)$ ($k\neq{i},
\ k\neq{j}$), $J(iiij)$, where
\begin{eqnarray}
&& J(ijkl)=\int{\int{\phi}^{*}}({\bf r}-{\bf R}_{i})\phi({\bf r}-{\bf R}_{k})
{e^{2}\over |{\bf r}-{\bf r}^{'}|}
\phi^{*}({\bf r}^{'}-{\bf R}_{j})\phi({\bf r}^{'}-{\bf R}_{l})
d{\bf r}d{\bf r}^{'}. 
\end{eqnarray}

The model Hamiltonian takes the following form~\cite{cmp}
\begin{eqnarray} \label{Ham}
H=&-&\mu \sum_{i\sigma}a_{i\sigma}^{+}a_{i\sigma}
+{\sum_{ij\sigma}}' (t_{ij}+nT_{1}(ij))a_{i\sigma}^{+}a_{j\sigma}
+{\sum_{ij\sigma}}'\left(T_2(ij)a_{i\sigma}^{+}a_{j\sigma}n_{i\bar{\sigma}}
+h.c.\right)
\nonumber\\ 
&+&U\sum_{i}n_{i\uparrow}n_{i\downarrow}
+{J\over 2}{\sum_{ij\sigma \sigma^{'}}}' a_{i\sigma}^{+}
a_{j\sigma{'}}^{+}a_{i\sigma{'}}a_{j\sigma},
\end{eqnarray}
where $a_{i\sigma}^{+}$, $a_{i\sigma}$ are creation
and destruction operators of electron on site $i$, 
$\sigma =\uparrow, \downarrow$, $n_{i\sigma}=a_{i\sigma}^{+}a_{i\sigma}$,
$n=\langle n_{i\uparrow}+n_{i\downarrow}\rangle$,
$\mu$ is the chemical potential, 
$t_{ij}$ is the band hopping integral of an electron 
from site $j$ to site $i$,
$U$ is the intra-atomic Coulomb repulsion 
and $J$ is the exchange integral for the nearest neighbours,
\begin{eqnarray*}
T_{1}(ij)=\sum_{\stackrel{k\neq{i}}{k\neq{j}}}J(ikjk)
\end{eqnarray*}
and $T_2(ij)=J(iiij)$ are the parameters of correlated hopping of electrons. 
The prime on the sums in Eq.~(\ref{Ham}) signifies that $i\neq{j}$. 

The peculiarities of the model described by the Hamiltonian~(\ref{Ham}) 
is taking into consideration the influence of the site occupation on the 
electron hoppings (correlated hopping), and the exchange integral.
In this model an electron hopping from one site to another is correlated 
(in contrast to similar generalized Hubbard models) both by the occupation
of the sites 
involved in the hopping process (with the hopping integral $T_2$) and the 
occupation of the nearest-neighbour sites (with the hopping integral $T_1$)
which we took into account by means of the Hartree-Fock approximation.

Thereby the correlated hopping, firstly, renormalizes the initial hopping 
integral (it becomes concentration- and spin-dependent) and, secondly,
leads to an independent on quasiimpulse shift of the band center, dependent on
magnetic ordering. Taking into account the quantity of second order $J$ 
is in principle necessary to describe ferromagnetism in this 
model.

To characterize the value of the correlated hopping we introduce dimensionless
parameters $\tau_1=\frac{T_1(ij)}{|t_{ij}|}$, 
$\tau_2=\frac{T_2(ij)}{|t_{ij}|}$.

\section{The Green function and energy spectrum in case of weak interaction}
The single-particle Green function satisfies the equation
\begin{eqnarray}
&&(E-\mu)\langle\langle a_{p\sigma}|a^{+}_{p'\sigma}\rangle\rangle_{E}
=\frac{\delta_{pp'}}
{2\pi}+\sum_{i}t_{ip}(n)\langle\langle a_{i\sigma}|a^{+}_{p'\sigma}
\rangle\rangle_{E}
\nonumber
\\ 
\nonumber
&&+\sum_{i}T_2(ip)\langle\langle a^{+}_{p\bar{\sigma}}
a_{p\bar{\sigma}}a_{i\sigma}|a^{+}_{p'\sigma}\rangle\rangle_{E}
+\sum_{i}T_2(ip)\langle\langle a^{+}_{i\bar{\sigma}}
a_{i\bar{\sigma}}a_{i\sigma}|a^{+}_{p'\sigma}\rangle\rangle_{E}
\\ 
\nonumber
&&+2\sum_{i}T_2(ip)\langle\langle a^{+}_{p\bar{\sigma}}
a_{i\bar{\sigma}}a_{p\sigma}|a^{+}_{p'\sigma}\rangle\rangle_{E}
+U\langle\langle n_{p\bar{\sigma}}a_{p\sigma}|a^{+}_{p'\sigma}\rangle
\rangle_{E}
\\ 
&&
+\sum_{j\sigma'} J
\langle\langle a^{+}_{j\sigma'}a_{p\sigma'}a_{j\sigma}
|a^{+}_{p'\sigma}\rangle\rangle_{E}.
\end{eqnarray}

Let us consider the system at weak intra-atomic Coulomb interaction 
($U<2w$, $w=z|t_{ij}|$ is the half of bare bandwidth, $z$ is the number of
nearest neighbours to a site). Then we can take into account the
electron-electron interactions in the Hartree-Fock approximation:
\begin{eqnarray}\label{appr}
\langle\langle a^{+}_{i\bar\sigma}a_{i\bar\sigma}a_{j\sigma}|a^{+}_{p'\sigma}
\rangle\rangle_{E}&\simeq&\langle a^{+}_{i\bar\sigma}a_{i\bar\sigma}\rangle
\langle\langle a_{j\sigma}|a^{+}_{p'\sigma}\rangle\rangle_{E};\\
\nonumber
\langle\langle a^{+}_{i\bar\sigma}a_{j\bar\sigma}a_{i\sigma}|a^{+}_{p'\sigma}
\rangle\rangle_{E}&\simeq&\langle a^{+}_{i\bar\sigma}a_{j\bar\sigma}\rangle
\langle\langle a_{i\sigma}|a^{+}_{p'\sigma}\rangle\rangle_{E}.
\end{eqnarray}
We assume that averages $\langle a^{+}_{i\sigma}a_{i\sigma}\rangle=n_\sigma$
are independent of the number of the site (the uniform distribution
of electronic density and magnetic moments is assumed). 

The approximation~(\ref{appr}) corresponds to the condition of weak 
electron-electron interaction, where doubly occupied states can be taken
into account by the effective mean field depending on the electron 
concentration and correlated hopping integral.

After the transition to Fourier representation we obtain for the Green 
function 
\begin{eqnarray}\label{grf}
&& \langle\langle a_{p\sigma}|a^{+}_{p'\sigma}\rangle\rangle_{\bf k}
=\frac{1}{2\pi}\frac{1}{E-E_\sigma(\bf k)},
\end{eqnarray}
where the energy spectrum is
\begin{eqnarray}\label{spectr}
&& E_\sigma({\bf k})=-\mu+\beta_\sigma+n_{\bar{\sigma}}U-zn_{\sigma}J+
t(n\sigma)
\gamma({\bf k});
\end{eqnarray}
here the spin-dependent shift of the band center is
\begin{eqnarray}
&& \beta_\sigma={2 \over N}\sum_{ij} T_2(ij) \langle 
a_{i\bar{\sigma}}^{+}a_{j\bar{\sigma}}\rangle,
\end{eqnarray}
$\gamma({\bf k})=
\sum\limits_{{\bf R}}e^{i{\bf kR}}$
(the sum goes over the nearest neighbours to a site)
and the spin and concentration dependent hopping integral is
\begin{eqnarray}\label{tef}
&& t(n\sigma)=t+nT_1+2n_{\bar{\sigma}}T_2+J \sum_{\sigma'}  \langle 
a_{i\sigma'}^{+}a_{j\sigma'}\rangle.
\end{eqnarray}

 From the energy spectrum~(\ref{spectr}) by neglecting the last term
in~(\ref{tef}) we obtain the result of papers~\cite{jps1,cmp}; 
neglecting the correlated hopping $T_{1}(ij)$ - the result of 
Ref.~\cite{amad}.

 The dependences of effective hopping integral on electron concentration 
and magnetization, a presence of the spin-dependent shift of band
center are the essential distinctions of single-particle energy spectrum
of the model described by Hamiltonian~(\ref{Ham}) from the spectrum of the 
Hubbard model in the case of weak interaction.

The concentration of electrons with spin $\sigma$ is 
\begin{eqnarray}
n_{\sigma}=\int\limits_{-\infty}^{+\infty}\rho(\epsilon)f(\epsilon)d\epsilon.
\end{eqnarray}
Here $\rho(\epsilon)$ is the density of states, $f(\epsilon)$ is the Fermi 
distribution function. Let us assume the rectangular density of states:
\begin{eqnarray}
\rho(\epsilon)=\frac{1}{N}\sum_{\bf k}\delta(\epsilon-\epsilon({\bf k}))=
\frac{1}{2w}\theta(\epsilon^{2}-w^{2}).
\end{eqnarray}

In the case of zero temperature we obtain:
\begin{eqnarray}
n_{\sigma}=\frac{\varepsilon_{\sigma}+w}{2w},
\end{eqnarray}
here the value
$\varepsilon_{\sigma}$ is the solution of the equation $E_{\sigma}(\varepsilon)=0$
from which we obtain
$\varepsilon_{\sigma}=\frac{\mu_{\sigma}}{\alpha_{\sigma}}$,
where
$\mu_{\sigma}=\mu-\beta_{\sigma}+zn_{\sigma}J-n_{\bar\sigma}U$
and
$\alpha_{\sigma}=1-\tau_1n-2\tau_2 n_{\bar\sigma}
-{zJ\over w}\sum\limits_{\sigma'}n_{\sigma'}(1-n_{\sigma'})$.

The system parameters are related by the equation
\begin{eqnarray}
\label{steq}
zJ+U+{\beta_{\downarrow}-\beta_{\uparrow}\over m}=2w\left(1-\tau_1n-\tau_2
-{zJ[n(2-n)-m^2] \over 2w}\right).
\end{eqnarray}

The shift of band center is obtained from
\begin{eqnarray}
\beta_{\sigma}=\frac{2}{N}\sum_{ij}T_2(ij)\langle a^{+}_{i \bar\sigma}
a_{j \bar\sigma} \rangle=-\tau_2 wn_{\bar\sigma}(n_{\bar\sigma}-1).
\end{eqnarray}

One can see that
\begin{eqnarray}
\label{beta}
\beta_{\downarrow}-\beta_{\uparrow}=2\tau_2 mw(1-n).
\end{eqnarray}
In particular, the concentration dependent shift~(\ref{beta}) 
of the centers of spin-up and spin-down electron bands
at $n<1$ is positive, at $n>1$ is negative.
In this manner we obtain the condition for the equilibrium value of 
magnetization:
\begin{eqnarray}
\label{steq}
1-\tau_1n-\tau_2(2-n)-{U \over 2w}={zJ \over 2w} [1+n(2-n)-m^2]. 
\end{eqnarray}
This result will be compared with result of next section.  
\section{Ground state energy}
To calculate the ground state energy of the model we use the formula
\begin{eqnarray}\label{E0} 
{E_0}={1\over 2N} \sum_{{\bf k}\sigma} \int\limits_{-\infty}^{+\infty}
(t_{\bf k}(n)+E)J^{\sigma}(E)dE.
\end{eqnarray}
Here
\begin{eqnarray} 
J^{\sigma}(E)={\delta (E-E^{\sigma}({\bf k})) \over 1+\exp{E-\mu\over kT}}
\end{eqnarray}
is the spectral intensity of Green function~(\ref{grf}). 
From Eq.~(\ref{E0}) one can obtain for the ground state energy the expression
\begin{eqnarray}\label{E1} 
E_0={1\over 2} \sum_{\sigma} \left(-\mu_\sigma n_\sigma 
-(1-\tau_1 n +\alpha^\sigma) n_{\sigma}(1-n_{\sigma})w \right),
\end{eqnarray}
which can be rewritten in the form 
\begin{eqnarray}\label{E2} 
&&E_0=E^{(0)}_0+E^{(2)}_0+E^{(4)}_0,\\
&&E^{(0)}_0={n\over 2}\left(-\mu+{U\over 2}n+{zJn(2-n)\over 4}
-(1-(\tau_1+\tau_2)n)(2-n)w\right), \nonumber \\
&&E^{(2)}_0=\left( 1-\tau_1 n-\tau_2 (2-n)-{U\over 2w}-
{zJ\over 4w}(1+n(2-n))\right){w\over 2}m^2, \nonumber \\
&&E^{(4)}_0={zJ\over 8}m^4. \nonumber 
\end{eqnarray}
One can see that the condition for equillibrium 
value of magnetization obtained  from expression $dE_0/dm=0$ 
is equivalent to Eq.~(\ref{steq}).
The position of the minimum of ground state energy depends on 
values of model parameters. In Fig.1 the energy difference between
paramagnetic and ferromagnetic states is plotted as a function of 
the magnetization. From this plot we can see that at some
values of the parameters a ferromagnetic ordering occurs with 
increase of inter-atomic exchange integral. The values of magnetization
at which the ground state energy of the system has a minimum depends
on magnitude of inter-atomic exchange.  
The condition of ferromagnetic ordering stability 
$d^2 E_0/dm^2<0$ can be obtained as 
\begin{eqnarray}
\label{umova}
{zJ \over 2w} >\frac{1-n\tau_1-\tau_2(2-n)
-{U \over 2w}}{1+n(2-n)}.
\end{eqnarray}
For the case of $zJ/2w=0$ from Eq.~(\ref{umova}) we obtain 
a generalization of the Stoner criterion which takes into account the 
correlated hopping
\begin{eqnarray}
U\rho(\epsilon_F) >1-n\tau_1-\tau_2(2-n).
\end{eqnarray}
From the condition of the minimum of ground state energy $dE_0/dm=
0$ one can also obtain for the value of exchange 
integral at fixed magnetization $m$ 
\begin{eqnarray}
\label{umova_m}
{zJ \over 2w} =
\frac{1-n\tau_1-\tau_2(2-n)-{U \over 2w}}{1+n(2-n)-m^2}.
\end{eqnarray}
The condition of full spin polarization ($m=n$) is
\begin{eqnarray}\label{fullsp}
{zJ \over 2w} >
\frac{1-n\tau_1-\tau_2(2-n)-{U\over 2w}}{1+2n(1-n)}.
\end{eqnarray}
From Eq.~(\ref{umova})-(\ref{fullsp}) one can conclude:
the competition between correlated hopping parameters $\tau_1$  and 
$\tau_2$  determines more favourable situation for the ferromagnetic ordering. 
For example, if $\tau_2> \tau_1$ then the systems with the electron 
concentration $n<1$ are more favourable to ferromagnetism than the systems 
with $n>1$ and vice versa (see Fig.~2,~3; the values above the line 
correspond to the ferromagnetic state of the system, below the line - 
to the paramagnetic one), but
the correlated hopping parameter $\tau_2$ influences the condition of 
ferromagnetic ordering stronger than the parameter $\tau_1$ 
(see Fig.~4). Note also that the region of partial spin polarization 
is narrowed with a deviation of electron concentration from half-filling 
(see Fig.~2,~5,~6).


One can see that in the less than half-filled band, correlated 
hopping leads to the stabilization of ferromagnetism as well as 
inter-atomic exchange interaction and intra-atomic Coulomb interaction;
the larger is the electron concentration $n$, the smaller is the  
critical value of the exchange integral for the occurrence of ferromagnetism.
These our results are in accordance with the results of 
Ref.~\cite{amad}. 

The magnetization found from the condition of minimum of the ground state 
energy is written as
\begin{eqnarray}
\label{m0} 
m={\left( {zJ \over 2w}\right)}^{-{1 \over 2}}
\left( {zJ \over 2w}(1+n(2-n))+\frac{U}{2w}+\tau_1 n+\tau_2 (2-n)-1
\right)^{ 1 \over 2},
\end{eqnarray}
The magnetization defined by Eq.~(\ref{m0}) is plotted as a function of 
$zJ/w$ for different values of $U/w$ in the Fig.7.
In the ferromagnetic state the magnetization rapidly increase with 
$zJ/w$ and reach its maximum value at relatively small values of
exchange integral what indicates the importance of inter-atomic exchange 
interaction for ferromagnetism of the system. The dependence of
the magnetization on concentration of electron is shown in Fig.8.
At fixed values of model parameters the change of electron concentration 
(doping of the system) can induce ferromagnetism.

In Fig.9 the dependence of paramagnetic and ferromagnetic ground state 
energies of the system on $zJ/w$  is plotted for different values 
of $n$. For all values of the electron concentration at sufficiently 
large values of $zJ/w$ the energy of the ferromagnetic state lies 
much lower then the paramagnetic one. The position of the transition point
depends on values of $U$ and $\tau_1$, $\tau_2$. With the increase of 
these parameters the critical value of inter-atomic exchange interaction
becomes smaller (the influence of the Coulomb repulsion is 
illustrated in Fig. 10).

The energy difference between the paramagnetic and ferromagnetic states, 
the value of magnetization as functions of band filling are plotted in Fig.11.
Depending on the value of $n$ the state of the system can be 
para- or ferromagnetic. For the zero values of correlated hopping parameters
(the curve 1 on Fig.11(a) ) in the system with increase of $n$ the ferromagnetic 
order occurs. Note that with further increase of $n$ the system becomes fully 
spin-polarized (see fig. 11(b) ) however in almost half-filled band the spin
polarization is partial again. In the more then half-filled band (in 
consequence of the electron-hole symmetry of this case) the noted 
magnetic behaviour repeats in inverse sequence. In the case of  
non-zero values of correlated hopping parameters (the curves 
2,3,4 on Fig.11(a) ) we have the valuable
decrease of ferromagnetic ground state energy comparing to paramagnetic one;
the influence of correlated hopping $\tau_1$ is stronger at $n>1$ 
and the influence of $\tau_2$ is stronger at $n<1$. In the case of  
equal values of $\tau_1$ and $\tau_2$ the electron-hole symmetry
retrieves (the curve 4 on Fig.11(a)). 

\section{Conclusions}

Although the consistent theory of ferromagnetic ordering in  transition
metal compounds can be constructed only in the model including the orbital 
degeneracy of the band, the qualitative character of observed properties 
can be interpreted in the framework of generalized model of non-degenerate
band. We have studied the ferromagnetism in this model of at weak intra-atomic 
interaction. The peculiarity of the studied model is taking into account 
besides the intra-atomic Coulomb  and inter-atomic exchange interactions 
the additional terms describing the interaction of the hopping electrons with 
other electrons (correlated hopping). In the considered case of weak 
electron-electron interaction the inter-atomic exchange interaction leads 
to the transition from a paramagnetic to a ferromagnetic state of the system.

The critical value of  exchange interaction strongly depends on model 
parameters. The correlated hopping causes the spin-dependent shift of the 
band center which can lead to
the ferromagnetism. This our result is in accordance 
with the results of Refs.~\cite{hir_t,amad}. In distinction from the model 
considered in Ref.~\cite{amad} our Hamiltonian contains additional mechanism of
correlated hopping which is shown to favour the ferromagnetic 
ordering. The intra-atomic Coulomb interaction is the factor favouring the
ferromagnetic ordering but this interaction itself 
can not stabilize the ferromagnetic ordering if $U<2w$ (see Fig.1).
The ground state energy has strong filling dependence and its behaviour 
can be changed by rise of the correlated hopping 
parameters. 

To conclude, in the case of weak electron-electron interaction 
the inter-atomic exchange interaction plays a substantial role in 
stabilization of the ferromagnetic ordering; 
the intra-atomic Coulomb interaction and    
the correlated hopping also favour ferromagnetism. The non-equivalence of 
less then half-filled and more than half-filled cases 
(concerning the condition of ferromagnetism realization)
is shown 
which is characteristic for the models with  correlated hopping of electrons.

\newpage

{\bf Figure captions}

Fig.1 The energy difference between paramagnetic and ferromagnetic states 
$E^{(m)} \over w$ as a function of the magnetization $m$ at ${U \over w}=1$, 
$\tau_1=\tau_2=0.2$ and $n=0.7$. Upper, middle and lower curves correspond 
to ${zJ \over w}=0 $, ${zJ \over w}=0.3$ and ${zJ \over w}=0.5$ respectively.

Fig.2 Ground state phase boundaries at ${U \over w}=0.8$, $\tau_1=0.15$ and 
$\tau_2=0.2$: curve 1 corresponds to the paramagnetic case, 
curves 2,3 - the cases of partially spin polarization, $50$ and 
$75$ percents respectively, curve 4 corresponds to the case of full spin 
polarization.

Fig.3 The critical value of the exchange integral as a function of 
$\tau=\tau_2$ at ${U \over w}=0.8$ and $\tau_1=0.15$:
curve 1 corresponds to $n=0.5$, curve 2 corresponds to $n=1$ and
curve 3 corresponds to $n=1.5$.

Fig.4 The critical value of the exchange integral as a function of electron 
concentration $n$ for given magnetization at $\tau_1=0.15$, $\tau_2=0.2$
(solid lines) and  $\tau_1=0.2$, $\tau_2=0.15$ (dashed lines),
${U \over w}=0.8$: curves 1,2,3 and 4 correspond to $m=0.1$, $m=0.5$,
$m=0.7$ and $m=0.9$ respectively.

Fig.5 Lines of constant magnetization for half-filled band: curves 1,2 
and 3 correspond to the cases $m=0$, $m=0.7$ and $m=1$ respectively.

Fig.6 Lines of constant magnetization for quarter-filled band: curves 1,2 and 3 
correspond to the cases $m=0$, $m=0.7$ and $m=1$ respectively.

Fig.7 The magnetization $m$ as a function of ${zJ \over w}$ 
at $\tau_1=\tau_2=0.1$ and $n=0.8$. Curves 1,~2 correspond to 
${U \over w}=1.2$ and ${U \over w}=0.7$ respectively.

Fig.8 The magnetization $m$ as a function of $n$ at ${U \over w}=1.2$ 
and $\tau_1=\tau_2=0.1$. Solid curve corresponds to ${zJ \over w}=0.3$, 
dashed curve corresponds to ${zJ \over w}=0.6$.

Fig.9 The paramagnetic (dashed line) and ferromagnetic (solid line)
ground state energies of the system as a function of ${zJ\over w}$ at 
${U \over w}=0.5$, $\tau_1=\tau_2=0.3$. Upper curves correspond to 
$n=1.2$, lower ones correspond to $n=0.7$.

Fig.10 The ground state energy as a function of ${zJ\over w}$ 
at $\tau_1=\tau_2=0.2$ and $n=0.8$. Curves 1,2 and 3 correspond to 
${U \over w}=1$, ${U \over w}=0.6$ and ${U \over w}=0.2$ respectively.

Fig.11 The energy difference between the paramagnetic and ferromagnetic 
states $E^{(m)} \over w$ (a) and the value of magnetization $m$ (b)  
as functions of band filling $n$ at ${U \over w}=1.2$, ${zJ\over w}=0.45$.
Curves 1,~2,~3 and~4 correspond to the cases $\tau_1=\tau_2=0$,
$\tau_1=0, \tau_2=0.05$, $\tau_1=0.05, \tau_2=0$ and $\tau_1=\tau_2=0.05$ 
respectively.
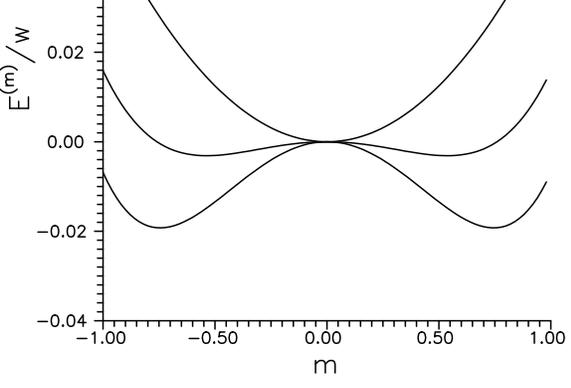
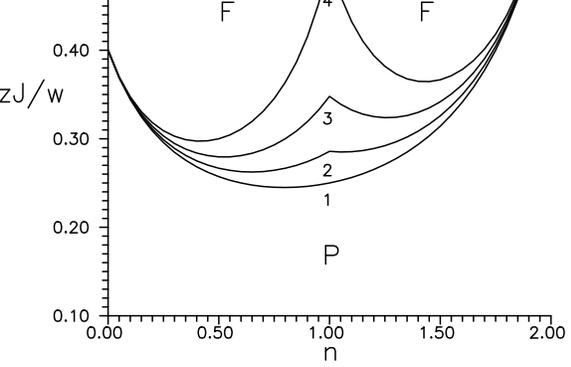
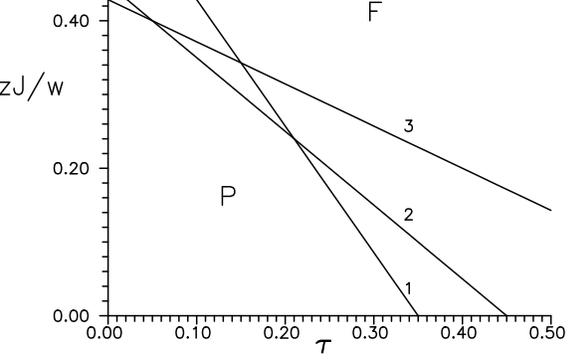
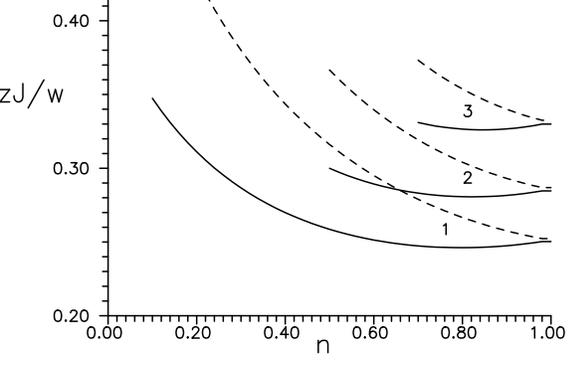
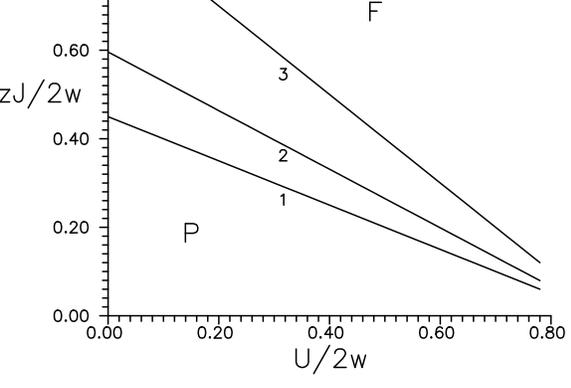
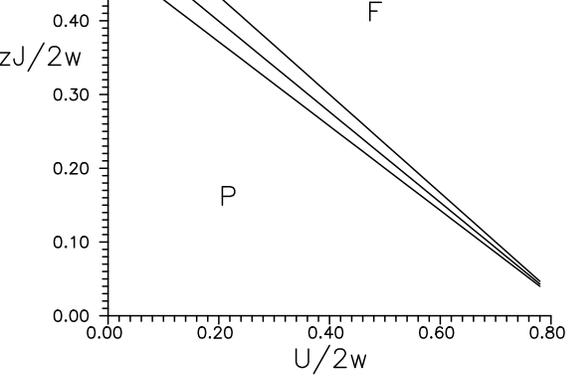
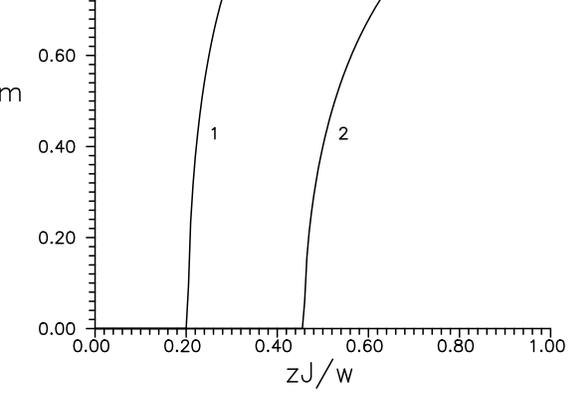
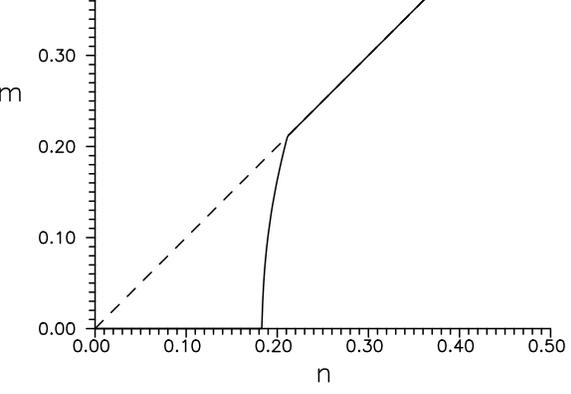
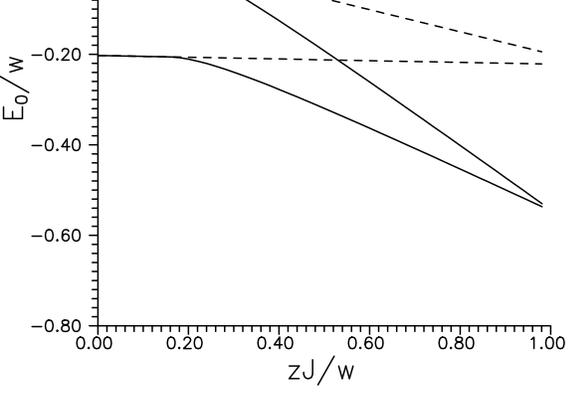
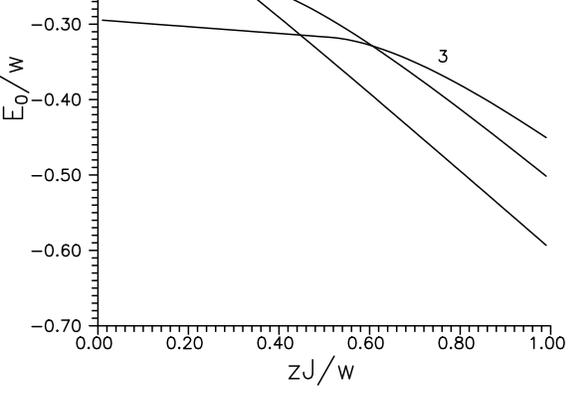
\end{document}